\documentclass[reprint,aps,prd,amsmath,amssymb,nofootinbib,superscriptaddress,floatfix]{revtex4-2}
\usepackage{bm}
\usepackage{times}
\usepackage{braket}
\usepackage{amssymb,amsbsy,amsmath,amsfonts}
\usepackage{graphicx}
\usepackage{float}
\usepackage{color}
\usepackage{morefloats}
\usepackage{rotating}
\usepackage{srcltx}
\usepackage{slashed}
\usepackage{subfigure}
\usepackage{multirow,diagbox}
\usepackage{verbatim}
\usepackage{tabularx}
\usepackage{tabularray}
\usepackage{epstopdf}
\usepackage[normalem]{ulem}
\usepackage{tikz}
\usepackage[unicode=true,colorlinks,linkcolor=blue,anchorcolor=blue,urlcolor=blue,citecolor=blue,breaklinks=true]{hyperref}

\newcommand{\imp}{\affiliation{State Key Laboratory of Heavy Ion Science and Technology, Institute of Modern Physics, Chinese Academy of Sciences, Lanzhou 730000, China}}

\newcommand{\snst}{\affiliation{School of Nuclear Science and Technology, University of Chinese Academy of Sciences, Beijing 101408, China}}

\newcommand{\qfnu}{\affiliation{College of Physics and Engineering, Qufu Normal University, Qufu 273165, China}}

\newcommand{\tju}{\affiliation{Center for Joint Quantum Studies and Department of Physics, School of Science, Tianjin University, Tianjin 300350, China}}

\newcommand{\scnt}{\affiliation{Southern Center for Nuclear-Science Theory (SCNT), Institute of Modern Physics, Chinese Academy of Sciences, Huizhou 516000, Guangdong, China}}

\begin{document}

\title{Study of the $e^+e^- \to \phi K^+K^-$ reaction within triangle dynamics and its implications for the $\phi(2170)$}

\author{Xiang Wei} \imp \snst
\author{Cheng Chen} \imp \snst
\author{Si-Wei Liu} \imp \snst
\author{Zu-Xin Cai} \qfnu \imp
\author{Gang Li} \email{gli@qfnu.edu.cn} \qfnu  
\author{Xiao-Hai Liu} \email{xiaohai.liu@tju.edu.cn} \tju
\author{Ju-Jun Xie} \email{xiejujun@impcas.ac.cn} \imp \snst \scnt

\begin{abstract}

We revisit the $e^+e^- \to \phi \pi^+\pi^-$ reaction within the $K_1$-$\bar{K}$-$K$ triangle dynamics framework, in which the $e^+e^-$ pair annihilates through one-photon exchange approximation to produce a $K_1\bar{K}$ pair, followed by the $K_1 \to\phi K$ decay and the final-state $K\bar{K} \to \pi^+\pi^-$ rescattering. With the same theoretical formalism and model parameters, the $e^+e^- \to \phi K^+K^-$ reaction is investigated, and it is found that the predicted total cross sections for the $e^+ e^- \to \phi K^+ K^-$ reaction are in good agreement with the existing BESIII measurements. Our study shows that the triangle singularity in the $\phi K^+K^-$ channel is strongly suppressed, because the higher $K^+K^-$ mass threshold shifts the kinematics away from the triangle singularity condition and the phase space near the $K_1\bar{K}$ threshold is very limited. Moreover, the interference between the tree-level and loop amplitudes eliminates the remaining signal. These combined effects naturally explain the absence of a distinct $\phi(2170)$ signal in the $e^+ e^- \to \phi K^+ K^-$ reaction, provide a strong test of the model, and reinforce the picture that both reactions are governed by the same underlying mechanism, in which the $\phi(2170)$ state is produced in the $e^+ e^-$ annihilation from the $K_1$-$\bar{K}$-$K$ triangle loop.
\end{abstract}

\date{\today}
\maketitle

\section{Introduction}\label{sec:introduction}

The $\phi(2170)$ state, previously known as $Y(2175)$, has attracted persistent experimental and theoretical attention since it was first observed by the \textit{BABAR} Collaboration in the $e^+e^- \to \gamma_{\mathrm{ISR}} \phi f_0(980)$ reaction~\cite{BaBar:2006gsq}. It has subsequently been observed in various processes by the Belle~\cite{Belle:2008kuo}, BES~\cite{BES:2007sqy}, BESIII~\cite{BESIII:2014ybv,BESIII:2017qkh,BESIII:2020gnc,BESIII:2021bjn,BESIII:2021aet}, and \textit{BABAR} Collaborations~\cite{BaBar:2007ptr,BaBar:2011btv}. However, a striking and intriguing feature has emerged from these experimental efforts: while a clear peak structure is consistently observed around $2.17~\mathrm{GeV}$ in the $e^+e^- \to \phi \pi^+\pi^-$ cross section~\cite{Belle:2008kuo,BaBar:2011btv,BESIII:2021aet}, no corresponding signal has been found in the $e^+ e^- \to \phi K^+ K^-$ reaction~\cite{BESIII:2019ebn}. The energy dependence of the $e^+ e^- \to \phi K^+ K^-$ cross sections differs significantly from that of the $e^+ e^- \to \phi \pi^+ \pi^-$ reaction. This pronounced channel-dependence poses a significant challenge for a conventional interpretation of the $\phi(2170)$ as an isolated resonance. 

On the theoretical side, many interpretations have been proposed for the inner structure of the $\phi(2170)$, which can be roughly classified into the following six categories: (I) a conventional excited $s\bar{s}$ state, interpreting it as a highly excited state of strangeonium, such as the $3^3S_1$ or $2^3D_1$ states~\cite{Ding:2007pc,Wang:2012wa,Afonin:2014nya,Pang:2019ttv,Zhao:2019syt,Li:2020xzs,Chen:2020xho}; (II) an $s\bar{s}g$ hybrid state~\cite{Ding:2006ya,Ho:2019org}; (III) a tetraquark state, including the compact $ss\bar{s}\bar{s}$ tetraquark interpretation based on QCD sum rules~\cite{Wang:2006ri,Chen:2008ej,Drenska:2008gr,Deng:2010zzd,Ke:2018evd,Agaev:2019coa,Liu:2020lpw}, as well as systematic studies of hidden-strangeness multiquark states within the dynamical diquark model~\cite{Jafarzade:2025qvx}; (IV) a $\Lambda\bar{\Lambda}(^3S_1)$ baryon-antibaryon bound state~\cite{Zhao:2013ffn,Deng:2013aca,Dong:2017rmg}; (V) an $S$-wave threshold effect~\cite{Chen:2008ej}; (VI) a hadronic molecular picture, which interprets $\phi(2170)$ as a $\phi f_0(980)$ molecule~\cite{MartinezTorres:2008gy,Gomez-Avila:2007pgn}, or as a $\phi K\bar{K}$ three-body system where the $K\bar{K}$ subsystem clusters as $f_0(980)$~\cite{Malabarba:2020grf,Malabarba:2023zez}. Although each theoretical model can accommodate the event concentration in specific regions, the observed channel-dependent behavior -- a prominent peak in $e^+ e^- \to \phi\pi^+\pi^-$ reaction and its complete absence in $e^+ e^- \to \phi K^+K^-$ reaction -- remains poorly explained by most of these theoretical frameworks. Furthermore, if the resonance structure observed in the $e^+ e^- \to K^+K^-\pi^0\pi^0$ reaction is interpreted as the $\phi(2170)$, none of the existing theoretical expectations agree well with the experimental results~\cite{BESIII:2020vtu}.

Recently, we proposed a novel interpretation of the $\phi(2170)$ signal observed in the $e^+e^- \to \phi \pi^+\pi^-$ reaction~\cite{Wei:2025ejv}, where we demonstrated that the peak structure around $2.17~\mathrm{GeV}$ can be naturally explained by the triangle singularity mechanism arising from the $K_1$-$\bar{K}$-$K$ intermediate state process, rather than by a genuine resonance. Within this picture, a pair of $K_1$ and $\bar{K}$ (or $\bar{K}_1$ and $K$) mesons is produced via one-photon exchange approximation, then the $K_1$ decays into $\phi K$, and the $K\bar{K}$ pair undergoes $S$-wave rescattering to produce $\pi^+\pi^-$, where the scalar meson $f_0(980)$ is dynamically generated~\cite{Oller:1997ti,Oller:1998zr,Gamermann:2006nm}. The triangle singularity develops precisely at the kinematic conditions corresponding to a center-of-mass energy around $2.17~\mathrm{GeV}$, giving rise to a pronounced enhancement that mimics a resonant signal. Moreover, by varying the mass of the intermediate $K_1$ meson, the resonance-like structure $R(2400)$ observed around $2.4~\mathrm{GeV}$ in the $e^+ e^- \to \phi\pi^+\pi^-$ reaction~\cite{BESIII:2021aet,Belle:2008kuo} can also be well reproduced within the same mechanism. To reproduce the available experimental data on the $e^+ e^- \to \phi \pi^+\pi^-$ reaction, it is found that two $K_1$ mesons with masses around $1613~\mathrm{MeV}$ and $1890~\mathrm{MeV}$ are needed~\cite{Wei:2025ejv}.

Along with this line, in this work, we adopt a more commonly used form factor and revisit the $e^+e^- \to \phi\pi^+\pi^-$ reaction within the same $K_1$-$\bar{K}$-$K$ triangle dynamics framework. Based on the fitted model parameters, we then extend our theoretical framework to the $e^+e^- \to \phi K^+K^-$ reaction. Within this framework, the higher threshold shifts the kinematic relations away from the region where the triangle singularity condition is fulfilled~\cite{Bayar:2016ftu}, meanwhile, the phase space near the $K_1 \bar{K}$ threshold is extremely limited, these two factors strongly suppress the triangle singularity, and then this suppression can provide a natural explanation for the experimentally observed absence of the $\phi(2170)$ peak in the $e^+ e^- \to \phi K^+K^-$ reaction, and would thus provide a direct test of the validity of the framework in a different process.

The remainder of this paper is organized as follows. In Sec.~\ref{formalism}, we present the effective Lagrangians, the scattering amplitudes for the $e^+e^- \to \phi\pi^+\pi^-$ and $e^+e^- \to \phi K^+K^-$ reactions, and the construction of the final-state rescattering amplitudes. In Sec.~\ref{results}, we show our numerical results for the fit to the experimental data, and present predictions for the $e^+ e^- \to \phi K^+ K^-$ reaction. Finally, a brief summary is given in Sec.~\ref{sec:summary}.

\section{Formalism} \label{formalism}

\begin{figure*}[htbp]
\centering
\includegraphics[trim=4.4cm 19.5cm 4cm 4.5cm, clip, width=1\textwidth]{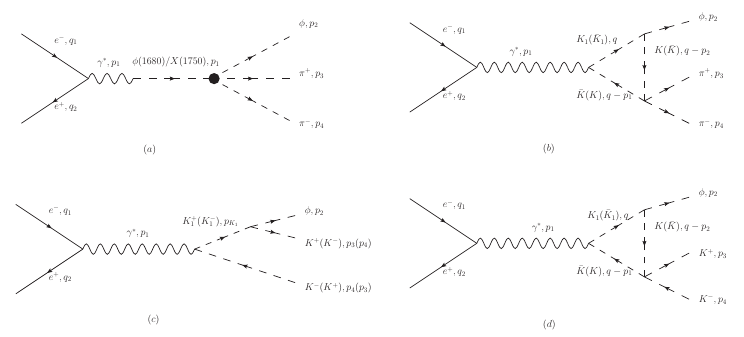}
\caption{Reaction mechanisms for the processes $e^+e^-\to\phi\pi^+\pi^-$ and $e^+e^-\to\phi K^+K^-$: (a) tree-level diagrams for $\phi\pi^+\pi^-$ production via the intermediate vector mesons $\phi(1680)$ and $X(1750)$~\cite{Wei:2025ejv}; (b) triangle loop diagrams for the $\phi\pi^+\pi^-$ final state; (c) tree-level diagram for $\phi K^+K^-$ production via an intermediate $K_1$ ($\bar{K}_1$) meson, followed by the $K_1(\bar{K}_1) \to\phi K(\bar{K})$ cascading decays; (d) triangle loop diagrams for the $\phi K^+K^-$ final state. The definitions of the kinematical variables ($q_1, q_2, p_1, p_2, p_3, p_4, p_{K_1}$) used in the present calculations are also shown.}\label{fig:process}
\end{figure*}

In Fig.~\ref{fig:process}, we show the tree-level diagrams and triangle loop diagrams for the $e^+e^- \to \phi \pi^+\pi^-$ reaction [Figs.~\ref{fig:process} (a) and (b)] and the $e^+e^- \to \phi K^+K^-$ reaction [Figs.~\ref{fig:process} (c) and (d)]. In the tree-level diagrams for the $e^+e^- \to \phi \pi^+\pi^-$ reaction [Fig.~\ref{fig:process} (a)], the virtual photon couples to the final state through the vector mesons $\phi(1680)$ and $X(1750)$. Meanwhile, the triangle loop mechanism is shown in Fig.~\ref{fig:process} (b), as done in Ref.~\cite{Wei:2025ejv}, where the scalar meson $f_0(980)$ is dynamically generated from the $S$-wave $K \bar{K}$ final-state interaction within the chiral unitary approach~\cite{Oller:1997ti,Oller:1998zr,Gamermann:2006nm}. In the tree-level diagrams for the $e^+e^- \to \phi K^+K^-$ reaction [Fig.~\ref{fig:process} (c)], $K_1 \bar{K}$ ($\bar{K}_1 K$) meson pairs are produced in the one-photon-exchange approximation, and $K_1$($\bar{K}_1$) subsequently decays into $\phi K^+$ ($\phi K^-$). The triangle-loop diagrams for the $e^+e^- \to \phi K^+K^-$ reaction [Fig.~\ref{fig:process} (d)] are analogous to those for the $\phi\pi^+\pi^-$ production. Note that we have also examined the contributions of the $\phi(1680)$ and $X(1750)$ mesons in the $e^+ e^- \to \phi K^+ K^-$ reaction, finding that their influence was negligible in the energy range of our focus since they are far from the mass threshold of $\phi K^+K^-$ system, their contributions are therefore not included in the final analysis.

Using the effective Lagrangian approach, we study the processes $e^+e^- \to \phi \pi^+\pi^-$ and $e^+e^- \to \phi K^+K^-$ through both tree-level diagrams and triangle loops. The interaction vertices involved in the diagrams of Fig.~\ref{fig:process} are described by the following effective interaction Lagrangians~\cite{Roca:2003uk,Roca:2004uc,Sakai:2019uig,Osipov:2020vad,Chen:2025xci}:
\begin{align}
\mathcal{L}_{\gamma R} &= -e\, g_{\gamma R} A^\mu R_\mu,\\
\mathcal{L}_{R\phi\pi\pi} &= g_{R\phi}\, R^\mu \phi_\mu \pi\pi,\\
\mathcal{L}_{\gamma K_1K} &= e\, g_{\gamma K_1\bar{K}} F^{\mu\nu}\partial_{\mu}K_{1\nu} K,\\
\mathcal{L}_{K_1\phi K} &= g_{K_1\phi K}\, K_1^\mu \phi_\mu K,
\end{align}
where $R$ denotes the vector meson field, representing either $\phi(1680)$ or $X(1750)$. $F^{\mu\nu}=\partial^\mu A^\nu -\partial^\nu A^\mu$ stands for the electromagnetic field
tensor. The coupling constants $g_{\gamma R}$, $g_{R\phi}$, $g_{\gamma K_1\bar{K}}$, and $g_{K_1\phi K}$ describe the strengths of the corresponding vertices.

With the effective Lagrangians given above, one can obtain the scattering amplitudes for $e^+e^- \to \phi\pi^+\pi^-$ and $e^+e^- \to \phi K^+K^-$ given in Fig.~\ref{fig:process} as follows, 

\begin{widetext}
\begin{eqnarray}
\mathcal{M}^{\pi\pi}_{\text{tree}} &=& \frac{-ie^2 g_R m_R^2 {T}_{\pi\pi \to \pi^+\pi^-}}{s(s-m_{R}^2+im_{R}\Gamma_{R})}\bar{v}(q_2,s_2)(\gamma^\alpha-\frac{\slashed{p}_{1}p_{1}^\alpha}{m_{R}^2}) u(q_1,s_1)\epsilon_\alpha(p_2,s_\phi) , \label{Mpipitree} 
\\
\mathcal{M}^{\pi\pi}_{\text{loop}} &=& \frac{i e^2 g_{K_1} \mathcal{F}(s)}{s}\int \frac{d^4q}{(2\pi)^4}    \frac{\bar{v}(q_2,s_2) (p_1\cdot q \gamma^\alpha - \slashed{q} p_1^{\alpha}) u(q_1,s_1)  \epsilon_\alpha (p_2,s_\phi) {T}_{K\bar{K} \to \pi^+\pi^-} }{(q^2-M_{K_1}^2 + i M_{K_1}\Gamma_{K_1})[(q-p_2)^2 - m_K^2] [(q-p_1)^2 - m_K^2] },\label{Mpipiloop}
\\
\mathcal{M}^{K\bar{K}}_{\text{tree}} &=& \frac{-ie^2 g_{K_1}\mathcal{F}(s)}{s(p_{K_1}^2-M_{K_1}^2+iM_{K_1}\Gamma_{K_1})}\bar{v}(q_2,s_2)(p_1\cdot p_{K_1}\gamma^\alpha-\slashed{p}_{K_1}p_{1}^\alpha) u(q_1,s_1)\epsilon_\alpha(p_2,s_\phi) , \label{MKKtree} 
\\
\mathcal{M}^{K\bar{K}}_{\text{loop}} &=& \frac{i e^2g_{K_1} \mathcal{F}(s)}{s}\int \frac{d^4q}{(2\pi)^4}    \frac{\bar{v}(q_2,s_2) (p_1\cdot q \gamma^\alpha -\slashed{q} p_1^{\alpha}) u(q_1,s_1)  \epsilon_\alpha (p_2,s_\phi) {T}_{K\bar{K} \to K^+K^-} }{(q^2-M_{K_1}^2 + i M_{K_1}\Gamma_{K_1})[(q-p_2)^2 - m_K^2] [(q-p_1)^2 - m_K^2] }.\label{MKKloop}
\end{eqnarray}
\end{widetext}
Here, $\bar{v}(q_2,s_2)$ and $u(q_1,s_1)$ are the Dirac spinors for the positron $e^+$ and electron $e^-$, respectively, while $\epsilon_\alpha(p_2,s_\phi)$ is the polarization vector of the $\phi$ meson, with $s_1$, $s_2$, and $s_\phi$ the corresponding spin polarization variables. The invariant mass squared of the initial $e^+e^-$ system is $s = p_1^2 = (q_1 + q_2)^2$. The symbols $M$ and $\Gamma$ with subscripts denote the masses and widths of the corresponding mesons. For the tree-level $\phi K^+K^-$ amplitude, the intermediate $K_1$ momentum $p_{K_1}$ is given by $p_{K_1}=p_2+p_3$ or $p_{K_1}=p_2+p_4$, as shown in Fig.\ref{fig:process}(c), corresponding to the two possible decay modes $K_1^+\to\phi K^+$ and $K_1^-\to\phi K^-$, respectively. Therefore, the tree-level amplitude for the $\phi K^+ K^-$ reaction is the sum of these two contributions.

Since the coupling constants $g_{\gamma K_1\bar{K}}$ and $g_{K_1\phi K}$ enter the scattering amplitudes through their product in the amplitudes and cannot be disentangled by the fit, we combine them into a single effective coupling, $g_{K_1} \equiv g_{\gamma K_1\bar{K}}\, g_{K_1\phi K}$, as shown in Eqs.~(\ref{Mpipiloop})-(\ref{MKKloop}). Motivated by the two distinct structures observed in the $\phi\pi^+\pi^-$ cross section, the $\phi(2170)$ peak and the $R(2400)$ bump, we include two $K_1$ states, labeled $K_1^a$ and $K_1^b$, which are assumed to be responsible for these two structures, respectively. Their effective couplings are denoted by $g_{K_1}^a$ and $g_{K_1}^b$, which will be determined by fitting them to the experimental data. Similarly, we combine $g_{\gamma R}$ and $g_{R\phi}$ into a single effective coupling, $g_R \equiv g_{\gamma R} g_{R\phi}$, with $R = \phi(1680)$ or $X(1750)$, as shown in Eq.~(\ref{Mpipitree}).

The $\mathcal{F}(s)$ in Eqs.~\eqref{Mpipiloop}-\eqref{MKKloop} denotes the form factor of the photon propagator, for which we adopt a commonly used dipole parametrization form~\cite{Drell:1971vx,Zhao:2006gw,Li:2007au,Chen:2024luh,Chen:2023oqs,Yan:2023nlb,Yan:2023yff,Chen:2025xci}:
\begin{equation}
\mathcal{F}(s)=\left(\frac{\Lambda^2}{\Lambda^2-s}\right)^2.
\label{formfactor}
\end{equation}
Here the $\Lambda$ is a cutoff parameter. Compared with the exponential form factor $e^{-b\sqrt{s}}$ used in our previous work~\cite{Wei:2025ejv}, the dipole form adopted in the present work exhibits almost the same $s$-dependence within the wide energy region of interest. It is worthy to mention that $\Lambda$ has positive value, hence $\mathcal{F}(s)$ will be divergent at $\Lambda^2 = s$. To evade this problem, one can restrict $\Lambda \textless m_{\phi} + 2m_\pi$. On the other hand, in the practical fitting process, since the cutoff parameter $\Lambda$ is strongly correlated with the coupling constants and also the widths and masses of the two $K_1$ states, we fix its value around 1 GeV in the fitting to avoid fit instability and reduce the number of free parameters.

\begin{figure}[htbp]
\centering
\includegraphics[scale=0.6]{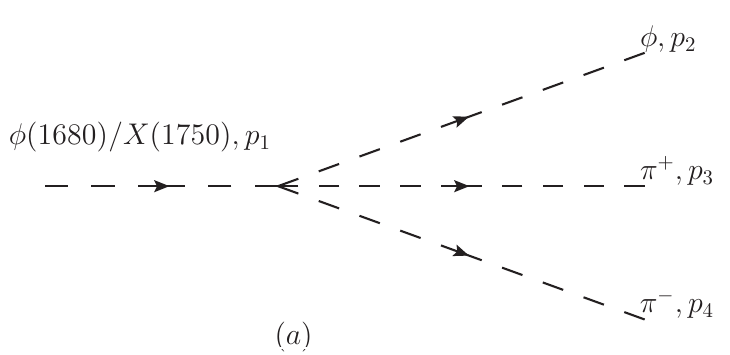}
\includegraphics[scale=0.6]{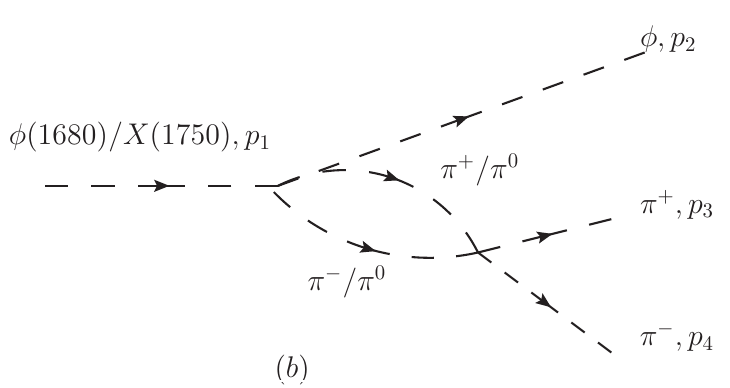}
\caption{Reaction mechanisms contributing to the $\pi^+ \pi^-$ final state: (a) direct production, (b) $\pi^+\pi^-$ and $\pi^0\pi^0$ rescattering.}
\label{fig:rescattering}
\end{figure}

In both the tree-level amplitude [Eq.~(\ref{Mpipitree})] and the loop amplitudes [Eqs.~(\ref{Mpipiloop}) and (\ref{MKKloop})], we take into account the $S$-wave final-state interactions described by the two-body rescattering amplitudes $T$ for the $\pi\pi$ and $K\bar{K}$ systems, respectively. For the $\pi^+\pi^-$ pair, the $G$-parity is $+1$, which restricts its isospin to even values; since the $\phi(1680)$ and $X(1750)$ resonances are isoscalars, isospin conservation selects the $I=0$ channel. The tree-level process in Fig.~\ref{fig:process}(a) then receives contributions from the three mechanisms as depicted in Fig.~\ref{fig:rescattering}. Using the isospin decomposition:
\begin{equation}
\ket{\pi\pi}_{I=0} = -\frac{1}{\sqrt{3}}\ket{\pi^+\pi^-} -\frac{1}{\sqrt{3}}\ket{\pi^-\pi^+} -\frac{1}{\sqrt{3}}\ket{\pi^0\pi^0},
\end{equation}
the full $\pi\pi \to \pi^+\pi^-$ rescattering amplitude is obtained as
\begin{align}
T_{\pi\pi\to\pi^+\pi^-} 
&= - \frac{1}{\sqrt{3}} \bigg( 2 + 2 G_{\pi\pi}\, t_{\pi^+\pi^-\to\pi^+\pi^-} \notag\\
&\qquad +  G_{\pi\pi}\, t_{\pi^0\pi^0\to\pi^+\pi^-} \bigg).
\end{align}
Here, $G_{\pi\pi}$ denotes the $\pi$-$\pi$ loop function of the $\pi\pi$ system, and $t_{ij}$ are the transition amplitude between the $i$-channel and $j$-channel, which can be obtained by solving the Bethe--Salpeter equation~\cite{Oller:1997ti,Oller:1998zr,Gamermann:2006nm}. $G_{\pi\pi}$ and $t_{ij}$ are dependent on the invariant mass of $\pi \pi$ system.

The same procedure applies to the $K\bar{K}$ rescattering in Fig.~\ref{fig:process}(b). For $I=0$, the isospin decomposition of the $K\bar{K}$ system is
\begin{equation}
\ket{K\bar{K}}_{I=0} = -\frac{1}{\sqrt{2}}\ket{K^+K^-} -\frac{1}{\sqrt{2}}\ket{K^0\bar{K}^0},
\end{equation}
from which corresponding rescattering amplitude is obtained as
\begin{equation}
T_{K\bar{K}\to\pi^+\pi^-} = - \frac{t_{K^+K^-\to\pi^+\pi^-} + t_{K^0\bar{K}^0\to\pi^+\pi^-}}{\sqrt{2}}.
\end{equation}
For the $K^+K^-$ final state, the pair does not have a definite $G$-parity, since it is a mixture of $I=0$ and $I=1$; the calculation is therefore performed directly in the charge basis without projecting onto isospin eigenstates.

It should also be noted that, as shown in Figs.~\ref{fig:process}(b) and \ref{fig:process}(d), the loop diagrams involve two equivalent charge-conjugate configurations, corresponding to $K_1\bar{K}$ and $\bar{K}_1 K$ production. Therefore, the total scattering amplitude can be expressed as:
\begin{align}
\mathcal{M}^{\pi\pi}_{\text{total}} &= \mathcal{M}_{\text{tree}}^{\pi\pi}(\phi(1680)) + \mathcal{M}_{\text{tree}}^{\pi\pi}(X(1750)) \notag\\
&\quad + 2\mathcal{M}_{\text{loop}}^{\pi\pi}(K_1^a) +2 \mathcal{M}_{\text{loop}}^{\pi\pi}(K_1^b), \label{Mpipitotal} \\
\mathcal{M}^{K\bar{K}}_{\text{total}} &= \mathcal{M}_{\text{tree}}^{K\bar{K}}(K_1^a,p_{K_1}=p_2+p_3)\notag\\
&\quad +\mathcal{M}_{\text{tree}}^{K\bar{K}}(K_1^a,p_{K_1}=p_2+p_4)\notag\\
&\quad + \mathcal{M}_{\text{tree}}^{K\bar{K}}(K_1^b,p_{K_1}=p_2+p_3)\notag\\
&\quad + \mathcal{M}_{\text{tree}}^{K\bar{K}}(K_1^b,p_{K_1}=p_2+p_4) \notag\\
&\quad + 2\mathcal{M}_{\text{loop}}^{K\bar{K}}(K_1^a) +2 \mathcal{M}_{\text{loop}}^{K\bar{K}}(K_1^b), \label{MKKtotal}
\end{align}
where the superscripts $a$ and $b$ denote the contributions from the two $K_1$ states. 

Then, the differential cross sections for the $e^+ e^- \to \phi \pi^+ \pi^-$ or  $e^+ e^- \to \phi K^+ K^-$ reactions can be straightforwardly written as:
\begin{align}
d\sigma_{\pi\pi(K \bar{K})} &= \frac{|p_\phi| |p^*_{\pi(K)}|}{16(\sqrt{s})^3(2\pi)^5} \sum_{s_1,s_2,s_\phi}|\mathcal{M}_{\text{total}}^{\pi\pi(K\bar{K})}|^2 \nonumber\\
&\quad \times dM_{\pi\pi(K\bar{K})} d\Omega_1 d\Omega^*_2,
\label{sigma}
\end{align}
where $M_{\pi\pi (K \bar{K})} = \sqrt{(p_3 + p_4)^2}$ is the invariant mass of the $\pi^+ \pi^-$ ($K^+ K^-$) system. $\Omega_1$ and $\Omega^*_2$ are the solid angles in the center-of-mass (CM) system of the $e^+ e^-$ collision and in the rest frame of the two-body $\pi^+ \pi^-$ ($K^+K^-$) final state, respectively. The $p_\phi$ and $p^*_\pi$ ($p^*_K$) are the three-momenta of the $\phi$ meson and $\pi^+$ or $\pi^-$ meson ($K^+$ or $K^-$ meson) in the CM frame and the $\pi^+\pi^-$ ($K^+K^-$) rest frame, respectively, which are given by
\begin{eqnarray}
&&|p_\phi|  = \frac{\sqrt{(s-(m_\phi + M_{\pi \pi(K\bar{K})})^2)(s-(m_\phi - M_{\pi \pi(K\bar{K})})^2)}}{2\sqrt{s}}, \nonumber \\
&&|p^*_{\pi(K)}| = \frac{\sqrt{M^2_{\pi\pi(K\bar{K})} -4m^2_{\pi(K)}}}{2}.  \nonumber
\end{eqnarray}

\section{Numerical results} \label{results}

In this work, we have eight parameters: (1) two effective coupling constants, $g_{\phi(1680)}$ and $g_{X(1750)}$, which combine the strengths of the vertices $\gamma$-$\phi(1680)$ and $\phi(1680)$-$\phi\pi\pi$, and of the vertices $\gamma$-$X(1750)$ and $X(1750)$-$\phi\pi\pi$, respectively, as depicted in Fig.~\ref{fig:process}(a); (2) two effective coupling constants, $g_{K_1}^a$ and $g_{K_1}^b$, which combine the strengths of the vertices $\gamma K_1\bar{K}$ and $K_1\phi K$ in Figs.~\ref{fig:process}(b)-(d), and correspond to the two $K_1$ states assumed to be responsible for the $\phi(2170)$ and $R(2400)$ structures, respectively; (3) the masses and widths of the two $K_1$ states, $M_{K_1}^a$, $M_{K_1}^b$, $\Gamma_{K_1}^a$, and $\Gamma_{K_1}^b$.

We perform a $\chi^2$ fit to the total cross section of the reaction $e^+e^- \to \phi \pi^+\pi^-$ measured by the BESIII~\cite{BESIII:2021aet} and \textit{BABAR}~\cite{BaBar:2011btv} Collaborations, comprising 57 data points below $\sqrt{s}=2600$~MeV. As discussed above, for $\Lambda=1$~GeV the $s$ dependence of the dipole form factor of Eq.~\eqref{formfactor} is almost identical to that of the exponential form factor used in our previous work~\cite{Wei:2025ejv} over the energy region of interest; we therefore fix the cutoff parameter at $\Lambda=1$ GeV. With this choice, the obtained $\chi^2/\mathrm{d.o.f.}$ is $2.5$. The fitted parameters are compiled in Table \ref{tab:resultsoffit}.

\begin{table}[htbp]
\renewcommand\arraystretch{1.5}
\centering
\caption{Free parameters fitted to the total cross sections of the reaction $e^+e^- \to \phi\pi^+\pi^-$ measured by the BESIII~\cite{BESIII:2021aet} and \textit{BABAR}~\cite{BaBar:2011btv} Collaborations.}
\begin{tabular}{c|c|c|c}
\hline   \hline
Parameter & Fitted results & Parameter & Fitted results \\
\hline
$g_{\phi(1680)}$ & $0.59 \pm 0.03$  & $g^a_{ K_1}$  & $-7.2 \pm 1.0 $\\
\hline
$g_{X(1750)}$  & $0.18 \pm 0.03$ & $g^b_{ K_1}$  & $-5.2 \pm 1.5$ \\
\hline
$M^a_{K_1}$ (MeV) & $1596 \pm 8$ & $\Gamma^a_{K_1}$ (MeV) & $76 \pm 22$ \\
\hline
$M^b_{K_1} $ (MeV)& $1886\pm 13$ & $\Gamma^b_{K_1}$ (MeV)  & $40 ^{+60}_{-40}$ \\
\hline\hline
\end{tabular}
\label{tab:resultsoffit}
\end{table}

\begin{figure}[htbp]
\centering
\includegraphics[trim=0cm 0.2cm 0cm 0cm, clip, width=0.47\textwidth]{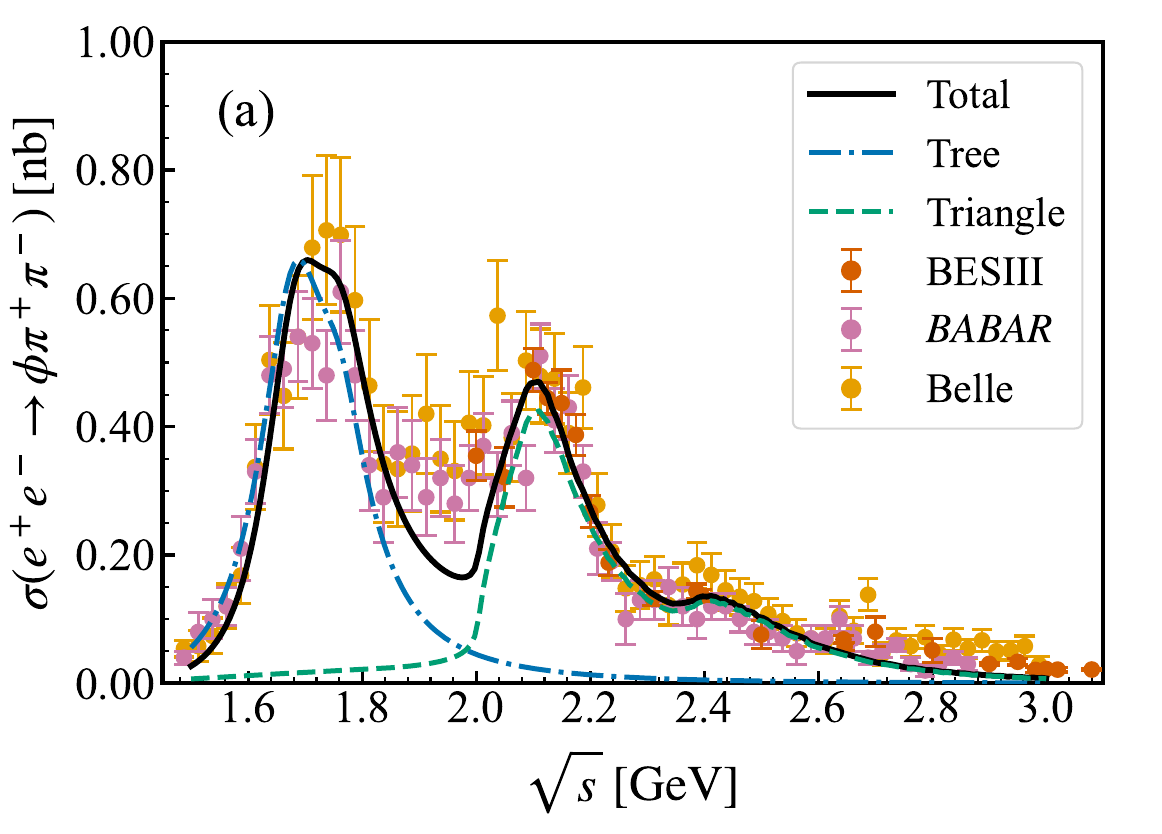}
\includegraphics[trim=0cm 0.4cm 0cm 0cm, clip, width=0.47\textwidth]{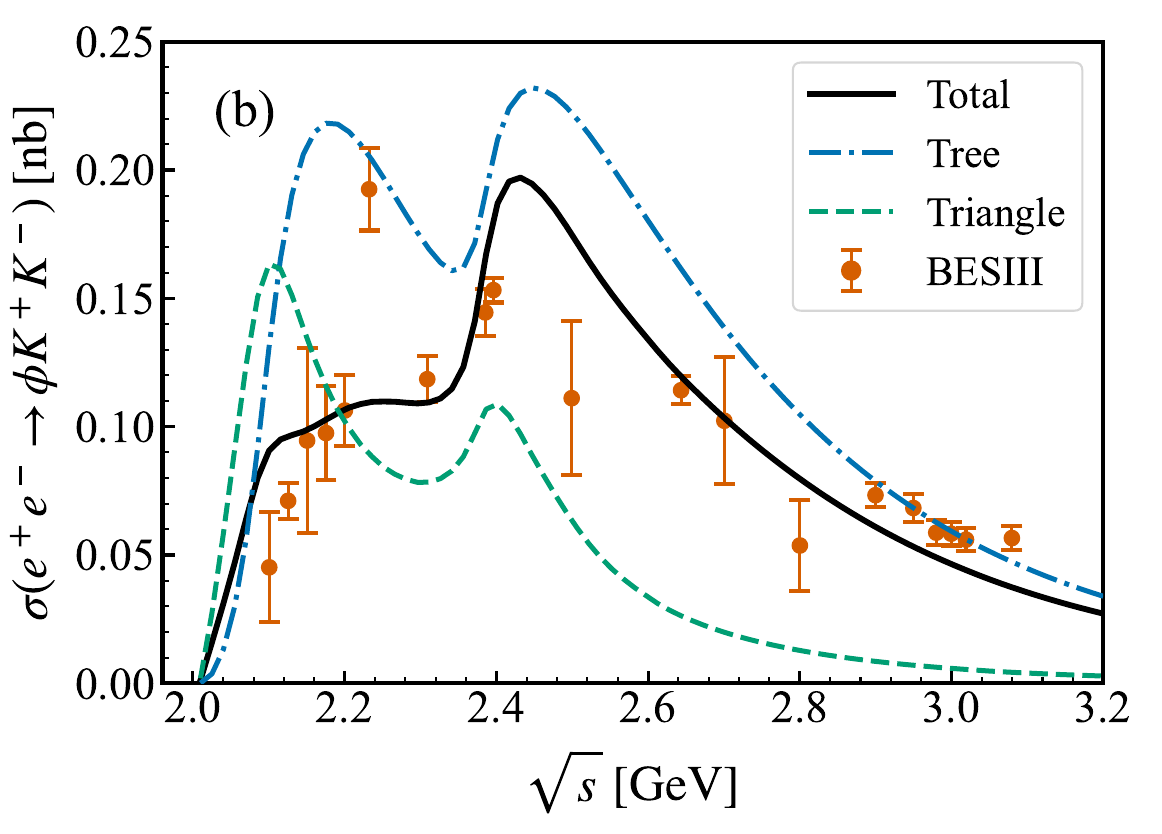}
\caption{Theoretical fitted results for the total cross sections of the $e^+e^- \to \phi \pi^+ \pi^-$ and $e^+ e^- \to \phi K^+ K^-$ reactions. The experimental data are taken from Refs.~\cite{Belle:2008kuo,BaBar:2011btv,BESIII:2021aet,BESIII:2019ebn}.}
\label{fig:Total}
\end{figure}

Using the above fitted values of the model parameters, we calculate the total cross sections of the $e^+ e^- \to \phi \pi^+ \pi^-$ and $e^+ e^- \to \phi K^+ K^-$ reactions, as shown in Fig.~\ref{fig:Total}. The theoretical results are compared with the experimental measurements from the Belle Collaboration~\cite{Belle:2008kuo}, BESIII Collaboration~\cite{BESIII:2021aet,BESIII:2019ebn} and \textit{BABAR} Collaboration~\cite{BaBar:2011btv}. Notably, we do not include the Belle Collaboration data~\cite{Belle:2008kuo} in our fitting procedure. As shown in Fig.~\ref{fig:Total}, the Belle measurements in the 2000--2100 MeV region lie systematically above the BESIII and \textit{BABAR} data, with relatively larger error bars. Including these data points would introduce a bias in the fit. The blue and red curves represent the contributions from tree diagrams [Fig.~\ref{fig:process}(a) and Fig.~\ref{fig:process}(c)], and triangle diagrams [Fig.~\ref{fig:process}(b) and Fig.~\ref{fig:process}(d)], respectively, while the black curve stands for the total contribution from all relevant diagrams.

From the fit results, we find that with the more commonly used dipole form factor, the fitted parameters are more reasonable and the uncertainties are substantially reduced compared with our previous work~\cite{Wei:2025ejv}. Although the fitted coupling constants change significantly, the main conclusions remain unchanged: the fitted masses and widths of the two $K_1$ mesons are $M_{K_1^a} = 1596 \pm 8$ MeV, $\Gamma_{K_1^a} = 76 \pm 22$ MeV, $M_{K_1^b} = 1886 \pm 13$ MeV, and $\Gamma_{K_1^b} = 40^{+60}_{-40}$ MeV,~\footnote{It is worth mentioning that, with the current $e^+ e^- \to \phi \pi^+ \pi^-$ experimental data, it is difficult to determine the widths of the two $K_1$ states, especially for the higher one.} which are fully consistent with our previous results within uncertainties. Furthermore, as shown in Fig.~\ref{fig:Total}(a), the $\phi(2170)$ peak and the $R(2400)$ bump structure are still well reproduced by the contributions from the $K_1$-$\bar{K}$-$K$ triangle loops of Fig.~\ref{fig:process}(b) and Fig.~\ref{fig:process}(d). 

For the $e^+ e^- \to \phi K^+K^-$ reaction, as shown in Fig.~\ref{fig:Total} (b), by incorporating a destructive interference between tree and triangle diagrams, our proposed reaction mechanism yields a predicted total cross section in good agreement with the experimental data reported by the BESIII Collaboration~\cite{BESIII:2019ebn} within uncertainties. Two bump structures of the $K_1$-$\bar{K}$-$K$ triangle loops associated with the tree level [Fig.~\ref{fig:process}(c)] contributions appear in the total cross section. Indeed, as we discussed in Ref.~\cite{Wei:2025ejv}, the triangle singularities in the diagrams of Fig.~\ref{fig:process}(b) appear when $K^+$ and $K^-$ in the triangle loop are on shell and the $\pi^+\pi^-$ invariant mass close to the mass of $f_0(980)$. This situation will be repeated in the diagrams of Fig.~\ref{fig:process}(c), since the $K^+$ and $K^-$ are external particles and, hence, they are on shell. And in this case the $K^+$ and $K^-$ propagators will be placed on shell and have their maximum contribution ((see also Ref.~\cite{Li:2025rlj}, where same conclusions were drawn for studying the $J/\psi \to \phi \eta \pi^0$ decay). However, the enhancement at $\sqrt{s} = 2.232$ GeV, very close to the $e^+ e^- \to \Lambda \bar{\Lambda}$ production threshold, can not be reproduced.

\begin{figure}[htbp]
\centering
\includegraphics[trim=1cm 1.3cm 0cm 0cm, clip, scale=0.42]{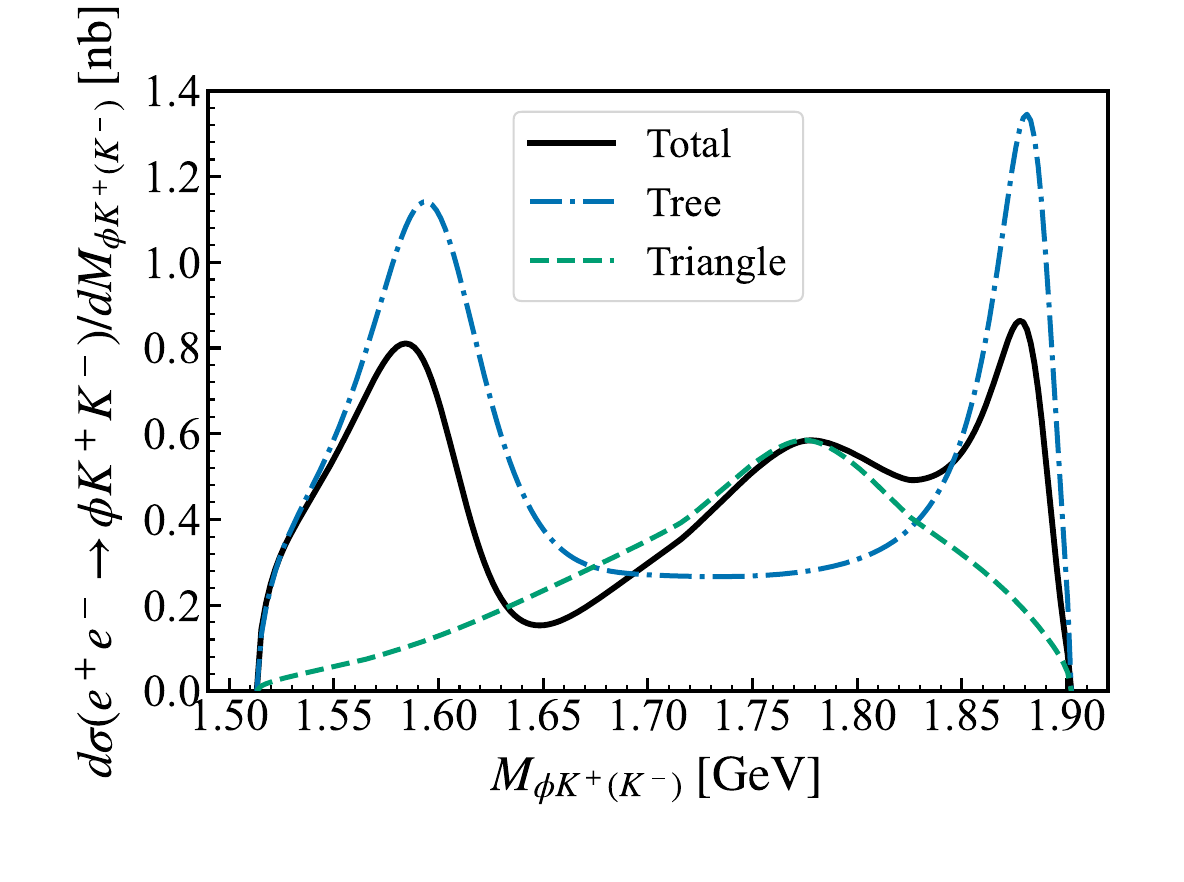}
\caption{Invariant mass distributions of $\phi K^+(K^-)$ systems in $e^+e^- \to \phi K^+K^-$ reaction at $\sqrt{s}=2.396$ GeV.}
\label{fig:phiK}
\end{figure}

As a further test of the model, we use the fitted parameters to predict the differential cross section of the $\phi K^+$ (or $\phi K^-$) invariant mass distribution at $\sqrt{s}=2.396$~GeV. The resulting $\phi K^+$ ($K^-$) invariant mass distributions are shown in Fig.~\ref{fig:phiK}. One can see that in addition to the two peak structures from the two $K_1$ mesons, there is also a bump structure around 1780 MeV, which is produced by the triangle loop mechanism. Since no experimental data exist for this observable at present, this prediction can serve as a benchmark for future measurements and offers an additional consistency check of the production of $\phi(2170)$ state in the $e^+ e^- \to \phi \pi^+ \pi^-$ reaction from the $K_1$-$\bar{K}$-$K$ triangle mechanism.

\section{Summary} \label{sec:summary}

The $\phi(2170)$ state has been observed in various hidden strange channels. However, its nature is a long-standing puzzle. In this work, we study the $\phi(2170)$ state in the $e^{+}e^{-} \to \phi K^{+}K^{-}$ reaction using effective Lagrangian approach, within the same $K_{1} - K - \bar{K}$ triangle dynamics framework previously applied to the $e^{+}e^{-} \to \phi \pi^{+}\pi^{-}$ channel~\cite{Wei:2025ejv}. Based on the one-photon exchange approximation, we model the production of a $K_{1}\bar{K}$ pair followed by the $K_{1} \to \phi K$ decay. The final-state $K\bar{K}$ rescattering is included, with the corresponding rescattering amplitudes obtained by solving the Bethe-Salpeter equation within the chiral unitary approach.

Our calculations show that in the $\phi K^{+}K^{-}$ final state, the increased threshold shifts the kinematics away from the triangle singularity conditions, while the extremely limited phase space near threshold further suppresses the singularity. Moreover, the strong interference between the triangle loop and tree-level amplitudes effectively cancels the remaining signal, so that no distinct $\phi(2170)$ structure is expected, consistent with experimental observations.

To determine the model parameters, we perform a fit to the total cross section data of $e^{+}e^{-} \to \phi \pi^{+}\pi^{-}$ from the BESIII~\cite{BESIII:2021aet} and \textit{BABAR}~\cite{BaBar:2011btv} Collaborations, achieving a reasonable fit quality of $\chi^{2}/\mathrm{d.o.f.} = 2.5$. Using these fitted parameters, we then compute the total cross section for $e^{+}e^{-} \to \phi K^{+}K^{-}$ and compare it with the BESIII measurements~\cite{BESIII:2019ebn}. The predicted results are in good overall agreement with the experimental data, reproducing the main trend as a function of $\sqrt{s}$, except for a few individual energy points. The determined masses and widths of the two $K_1$ mesons are:
\begin{align}
    M_{K_{1}^{a}} &= 1596 \pm 8 \text{ MeV}, &\Gamma_{K_{1}^{a}} &= 76 \pm 22 \text{ MeV},\\
    M_{K_{1}^{b}} &= 1886 \pm 13 \text{ MeV}, &\Gamma_{K_{1}^{b}} &= 40 ^{+60}_{-40} \text{ MeV}.
\end{align}

In conclusion, our results demonstrate that the $\phi(2170)$ peak observed in the $\phi \pi^{+}\pi^{-}$ channel, together with its absence in the $\phi K^{+}K^{-}$ channel, can be consistently explained within the same $K_{1} K \bar{K}$ triangle dynamics. This reinforces the picture that the $\phi(2170)$ structure is more likely a dynamical effect arising from the triangle singularity and its pole contributions, rather than a conventional isolated resonance. It is expected that more experimental information will be available in the near future~\cite{BESIII:2023dhc}, which will greatly improve our knowledge of the production mechanism of the $e^+ e^- \to \phi K^+ K^-$ reaction.

\begin{acknowledgments}

We would like to thank Prof. Wen-Biao Yan for useful discussions. This work is partly supported by the National Key R\&D Program of China under Grant No. 2023YFA1606703; by the National Natural Science Foundation of China under Grant Nos. 12575094, 12435007, 12475081, 12235018, and 12361141819; and by Taishan Scholar Project of Shandong Province under Grant No. tsqn202607074.

\end{acknowledgments}

\normalem
\bibliography{reference.bib}

\end{document}